\documentclass[fleqn]{article}
\usepackage{amscd,amsmath,amssymb}
\usepackage{emlines2,bezier,amscd,amsmath}
\usepackage[dvips]{graphicx}
\title{Are different geometries really that different?}
\author{Sergey  S. Kokarev\thanks{logos-distant@mail.ru}}
\date{Research Institute of Hypercomplex Systems in Geometry and Physics, RSEC "Logos"\,
                       Yaroslavl}

\begin{document}

\maketitle

\begin{abstract}
Here is presented a concept of centrogeometry which can be seen as a
combination of the concept of point-like observer with an idea of Poincar\'{e}'s that different geometries are principally equivalent.
As it is to be shown later, all centrogeometries are obtained from each other by general
deformation (i.e. active coordinate transformations). Isometries of centrogeometries are equivalent to those of the Euclidean
centrogeometry as described by common diffeomorphisms of the Euclidean spheres. There are discussed physical aspects of
centrogeometry in the context of chronogeometry, mechanics and cosmology.
\end{abstract}

\section{Introduction}

Since the creation of General Relativity theory the principle of geometrization of physical laws became widely applied everywhere \cite{vlad,kok}.
Basically it enables us to abandon, partially or fully, the language of forces and interactions within manifolds that, in place of  Euclidean geometry
structures,
have some more general structures such as generalized connection, curvature, extra dimensions, bundle structure and so on \cite{kob}.
Principle of geometrization is an essential part of modern "theories of everything" such as superstring theory and M-theory \cite{green}.

In the process of implementation of this principle the meaning of what we call geometry widens. Unlike the standard school definitions
according to which geometry is seen as a part of mathematics studying properties of points, lines, surfaces and their relations,
modern geometry studies relations between abstract objects. Some of these do not have even a remotest prototype in Euclidean geometry.

Under such general understanding of geometry the principle of geometrization loses its restrictive role that a physical principle
should have and becomes a general mode of reasoning we apply to the world. We "see"\, the world through the "lenses"\, of our geometries and describe
it in terms of our geometrical definitions. Different geometries play in this case the same role as different reference
frames do in solving problems of standard mathematical physics. When solving these problems, we do not interest ourselves as to which of
the reference frames is more correct. Moreover, according to coordinate description all reference frames are equal. There remains only
the question as to what possible reference frame would be more convenient and describe the phenomena considered in the easiest and most
adequate way. Following the same analogy between geometries and reference frames we can arrive at the same conclusion which was already
stated in 1909  by A. Poincar\'{e}, albeit in somewhat milder form: {\it all geometries are equal in a sense that any can be used
to describe any physical phenomena.} But different phenomena classes would be a described differently in different
geometries.
Some geometries would give an easy, elegant and symmetric description, in others it will look much more complex and complicated.
For example, we could explain all classical experiments in GR (displacement of Mercury's perihelion, bending of light rays,
red shift etc.) within the scope of Newtonian physics, modifying Newton's gravity law or even by means of introduction of new forces.
But, based on Riemannian geometry and its inherent notions, GR can describe these effects and foretell quite a few others without extra
modifications and without introduction of essences foreign to this theory. Thus, {\it simplification of description by means of substitution
of one geometrical paradigm by another can serve as a clue to the geometrization principle
development.}

Differential geometry allows us to study those properties of geometrical objects that do not depend on reference frames. Similarly, we can put
up a question: {\it are there properties of physical objects (maybe understood in a wider or abstract sense) that would not depend on the choice of
geometry?}
To answer this question we'll have to use a rather general approach to different physical theories where different geometries would play the role
of reference frames and where laws and equations would be formulated in terms of some above-geometrical metalanguage,
invariant as regards to switching of geometries. A more or less detailed version of such an approach does not exist as yet.

Another complex of ideas brought about by Poincar\'{e} in the work cited above is related to the nature of geometrical notions and geometry as a
whole. Having analyzed our geometrical notions and their invariants, Poincar\'e comes to the conclusion that they're an organic part
of subject's --- i. e.  human ---  perception and complexes of sensations (mostly visual and motive ones). Poincar\'{e}
was among the first who drew attention to the fact that, although not strictly deductions from logical analysis of our sensations,
geometrical notions rely heavily on them. Being abstractions from complicated and correlated complexes of primary visual, motive,
auditory, haptic etc. sensations, these notions are objects of the mind's "upper storey".
The role that complexes of sensations play in forming of physical notions was discussed by Mach \cite{mach},
but in a wider context that is the basis for philosophical {\it empiricism.} A more up-to-date overview of correlations between perception and physics
 can be found in a special appendix in \cite{bohm}.  The work \cite{kok1} offers a brief study of the visual perceptive space
 complete with a formal scheme in which physical geometry and physical laws are being deduced from complexes of sensations.

In present paper we formulate a concept of "centrogeometry"\, in which both ideas of Poincar\'{e}'s
are united. On the one hand the concept illustrates the "plasticity"\, of metric geometry in its appropriate
formulation, on the other hand it describes the physical concept of point-like observer.
It will be shown that for such point-like  observer all metric geometries are equivalent and the equivalency is accurate within arbitrary deformation
(gauge) of coordinate marks. "Isometries"\, of such centered geometries (linear in one gauge, but turning nonlinear in another one)
prove to be equivalent too. These isometries do not belong to classical Lie isometries of metric
manifolds because basically they are not related to two-point metric, but to a one-point metric.
Some physical applications of centrogeometry will be discussed in the Conclusion.

\section{General construction}

Let us remind that {\it metric} $\rho$  on a set $\mathcal{M}$ is a mapping of the kind: $\mathcal{M}\times\mathcal{M}\to R_+,$
that satisfies the following conditions:
\begin{enumerate}
\item
$\rho(p,p)=0;$
\item
$\rho(p,q)=\rho(q,p);$
\item
$\rho(p,q)\le\rho(p,r)+\rho(r,q)$
\end{enumerate}
for any $p,q,r\in\mathcal{M}.$ The value $\rho(p,q)$ is called {\it distance between points $p$ and
$q.$} The set $\mathcal{M}$ with some metric $\rho$ is called {\it metric space.}
For example, Riemannian manifolds are metric space, because their metrics $\rho$, at least locally, can be calculated according to formula:
\[
\rho(p,q)\equiv\text{length}[\Gamma_{pq}],
\]
where $\Gamma_{pq}$ is a geodesic line that joins $p$ and $q$ and is defined through standard equation of the
geodesics.
On the other hand, any metric manifold can always
be viewed as a Riemannian one if we define the norm of vector $X$ at the point with  coordinates $x$ according to the formula:
\[
|X|\equiv X^\alpha\left.\frac{\partial}{\partial
y^\alpha}\rho(x,y)\right|_{y=x+X}.
\]

Let us fix a point $p$ on a metric manifold $\mathcal{M}$ and consider value $\rho_p(x)\equiv\rho(p,x).$ Let us call the function $\rho_p(x)$
{\it  a centrometric at the point $p$.} Centrometric determines a system of geometrical relations in the vicinity
of point $p$ which we would call {\it centrogeometry} associated with the said point. As it will be shown later, centrogeometry is a substantially
less rigid construction than geometry based on two-pointed metric. We've set centrometric according to a certain metric.
It is apparent, however, that initially the construction of centrometrics can be defined on a manifold with a picked point that has no metric.

Let us consider diffeomorphisms  $\phi:\ \mathcal{M}\to \mathcal{M}$ and $f:\ R\to
R,$  that define some new function:
\begin{equation}\label{deforma}
\rho'_p\equiv f\circ\rho\circ\phi:\ R_+\to R_+,
\end{equation}
satisfying general conditions for metric and leave point $p$ unmovable.
From the active point of view it can be said that map $\phi$ determines
the manifold deformation, and map $f$  determines the deformation of the length scales.
Then it is natural to call the function $\rho'_p$  {\it deformed
centrometric.}

In majority situations centrometric $\rho_p$ possesses the following remarkable property: it has star-like
in relation to point $p$  surfaces of level (metric spheres),
i.e. in a some coordinate system on $\mathcal{M}$ centered  at the  point $p$  all rays of the kind $\lambda x$ under
$\lambda>0$ and  $x$ lying in a certain vicinity of point $p,$ intersect any sphere $\rho_p=\text{const}$  exactly at one point.
Moreover, typically in this coordinate system function $\rho_p(\lambda x)$ is a monotonically  increasing function of parameter $\lambda$
in a certain vicinity of the point
$p.$ In view of specific role of the point $p$ it is natural to restrict deformations $(f,\phi$) by
the following additional requirement: $\phi$ preserve star-like character of the metric spheres
$\rho=\text{const}$ and $f$ is isotonic mapping, i.e. $f$ preserve ordering of fibers $\rho=\text{const}.$
We shall denote $\Xi(\mathcal{M},p)=\{\rho_p\}$ class centrometrics
with star-like metric spheres and $\text{Diff}_p$ --- deformations of
the form (\ref{deforma}), preserving star-likeness property of
centrometrics and ordering of their metric spheres.

We are going to show, that all centrometrics $\rho_p$ satisfying
some very general conditions, are connected to each other by some
deformation.

\bigskip

{\bf THEOREM (on deformational equivalence)} {\sf For any pair $(\rho_1,\rho_2)$ of centrometrics of class $\Xi(\mathcal{M},p)$
there will be a pair of diffeomorphisms $(\phi,f)$ from $\text{Diff}_p$, so that
\begin{equation}\label{deform}
\rho_1=f\circ\rho_2\circ\phi.
\end{equation}
in a certain vicinity of point $p.$}

\begin{figure}[htb]
\centering \unitlength=0.50mm \special{em:linewidth 0.4pt}
\linethickness{0.4pt} \footnotesize \unitlength=0.70mm \special{em:linewidth
0.4pt} \linethickness{0.4pt}
\unitlength=1.00mm
\special{em:linewidth 0.4pt}
\linethickness{0.4pt}
\begin{picture}(33.00,44.50)
\put(18.83,20.00){\circle*{1.37}}
\bezier{128}(12.17,24.83)(5.33,8.50)(18.33,13.83)
\bezier{64}(12.17,24.83)(15.00,32.17)(23.17,31.00)
\bezier{72}(23.17,31.00)(31.50,27.83)(26.83,20.00)
\bezier{44}(26.83,20.00)(22.50,15.50)(18.33,13.83)
\bezier{144}(7.17,26.17)(5.67,44.50)(20.17,35.33)
\bezier{88}(20.00,10.50)(10.67,15.17)(7.17,26.17)
\bezier{88}(20.17,35.33)(28.67,28.00)(30.83,17.00)
\bezier{120}(20.00,10.50)(33.00,2.17)(31.00,17.00)
\emline{18.83}{20.00}{1}{16.50}{37.33}{2}
\emline{18.83}{20.00}{3}{31.33}{14.17}{4}
\emline{18.83}{20.00}{5}{10.50}{13.83}{6}
\put(17.50,30.50){\circle*{1.00}}
\put(16.50,37.33){\circle*{1.00}}
\put(24.17,17.50){\circle*{1.05}}
\put(31.17,14.33){\circle*{1.05}}
\put(13.00,15.67){\circle*{1.05}}
\put(10.33,13.83){\circle*{1.00}}
\emline{16.33}{30.83}{7}{15.67}{36.83}{8}
\emline{15.67}{36.83}{9}{16.33}{34.33}{10}
\emline{16.33}{34.33}{11}{16.33}{34.33}{12}
\emline{15.50}{34.50}{13}{15.67}{37.00}{14}
\emline{25.67}{17.67}{15}{30.50}{15.17}{16}
\emline{30.50}{15.17}{17}{28.50}{16.83}{18}
\emline{28.33}{15.83}{19}{30.50}{15.17}{20}
\emline{10.67}{14.67}{21}{12.00}{15.83}{22}
\emline{12.00}{15.83}{23}{10.33}{15.17}{24}
\emline{11.17}{14.50}{25}{12.00}{16.00}{26}
\put(20.00,20.50){\makebox(0,0)[lc]{$p$}}
\put(15.33,30.56){\makebox(0,0)[rb]{$\rho_1=\text{const}$}}
\put(10.89,39.00){\makebox(0,0)[cb]{$\rho_2=\text{const}$}}
\end{picture}
\caption{\small Mapping of metric starlike spheres in different centrometrics}\label{diff1}
\end{figure}
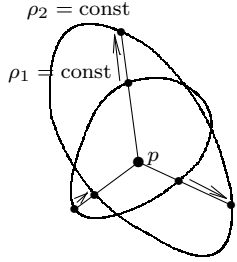

{\it Proof.}
First let us establish that in a certain vicinity of $p$ there is a diffeomorfism $\phi,$
that transforms spheres of $\rho_1$  metric into spheres of $\rho_2$ metric.
In order to do this we need to switch from coordinates $x$ given in the theorem conditions to  a spheric
(in Euclidean sense) reference frame $(R,\theta)$ with center in point $p,$ where $R$ is Euclidean length,  $\theta$ is set of $n-1$
spherical angles ($n=\dim\mathcal{M}).$ Metric spheres of $\rho_1$ and $\rho_2$   metrics in starlikeness  vicinity are defined by equations of the
kind: $R=\varphi_1(\theta)$
and $R=\varphi_2(\theta)$ and respectively. Mapping   $\phi$ in this vicinity can be defined as follows:
\[
\phi:\ (R,\theta)\mapsto(R',\theta),\ \text{где}\ \rho_1(R,\theta)=\rho_2(R',\theta).
\]
Fig. \ref{diff1} illustrates this definition.
In view of the starlike nature of the spheres, smoothness of metrics and equanimity between argumentations  $\rho_1$ and $\rho_2$
this map can be defined as diffeomorphism. On the other side, construction of this map represents spheres of $\rho_1$  metrics in spheres of $\rho_2$
metrics with the same radius. That means $\rho_1=\rho_2\circ\phi.$

Construction given above shows that related deformation allows for ambiguous definitions.
First option is compositions with arbitrary  smooth slips along spheres $\rho_1.$
Second option is monotonous mapping of spheres with different radius.
In latter case we would need a compensatory function $f$ that would "straighten"\, the map into identical one.
Apparently it always exists. If we choose to ignore smoothness and bijectivity of deformations,
conditions of the theorem could be substantially loosened, but in present paper there's no need for it. The theorem is
proven.$\Box$

Let us assume that $\chi$ is  the centrometric $\rho_p$ isometry where point $p$ is fixed:
\[
\rho_p\circ\chi=\rho_p.
\]
Let's then consider the deformed centrometrics $\rho'_p=\rho_p\circ\phi.$ Due to map bijectivity we obtain a series of equalities:
\[
\rho'_p\circ(\phi^{-1}\circ\chi\circ\phi)=\rho_p\circ\phi\circ\phi^{-1}\circ\chi\circ\phi=\rho_p\circ\phi=\rho'_p,
\]
from which it follows that the map:
\begin{equation}\label{deff}
\chi'=\phi^{-1}\circ\chi\circ\phi
\end{equation}
is an  isometry of the deformed metric $\rho'_p,$ which can be called {\it isometry deformed along $\phi$}.
In general case, deformed isometry is a nonlinear mapping of the initial isometry $\chi,$ which in its turn would be wider than classical
 Lie's isometry of manifold $\mathcal{M}$ with fixed point $p.$
In view of deformational connectedness of centrometrics of class $\Xi(\mathcal{M},p),$ we can draw a conclusion regarding deformational
connectedness of all centrometrics isometries: {\it any two isometries of a class $\Xi(\mathcal{M},p)$ are connected to each other by
deformation
according to formula (\ref{deff}).}

And that's not all there's to it. Let us consider Euclidean centrogeometry in a spheric reference frame.
In these coordinates centrometrics will acquire the following form: $\rho_E(r,\theta)=r,$ where $\theta$ is a family of $n-1$ spherical angles.
It is apparent that the isometry group of this centrometrics is in fact transformations that invert metric spheres into themselves.
These  transformations can be described by smooth mapping $\Phi:\ R^{n-1}\to R^{n-1}$ of the kind:
\begin{equation}\label{gizo}
\theta'=\Phi(\theta).
\end{equation}
Unlike classical isometries of metrics, these are infinite-dimensional transformations of the type $\infty^{n-1}.$
General isometry group of any centrometrics obtained by Euclidean metrics deformation has the form: $\Phi'=\phi^{-1}\circ\Phi\circ\phi$
and descends from Euclidean isometry group.

The properties of centrometrics established here allow us to maintain that, unlike metrics which of course cannot be inverted into each other
by manifold deformation (with the exception of isometrical metrics), all centrometrics are equivalent in a sense as mentioned before. Unlike metrics,
centrometrics appear to be a "cruder"\, object since it is attached to one selected point of the manifold.
The theorem we've proven means that by setting centrometrics we will not fix a certain geometry of manifold in a classical sense of the word,
but set (locally) a whole class of centrogeometries, one of which is, for example, Euclidean centrogeometry.
It looks pretty arbitrary, but nevertheless, in some respect concept of centrometrics is better at reproduction of relations between geometry
and experience than standard two-point metrics. In particular this conception ideally suits the physical concept of "point-like"\,
observer (see section 5).

\section{Examples}

In this section we'll examine a few specific examples illustrating general construction of the previous section.

\subsection{ Centrogeometry of Euclidean  plane and sphere}

Differential geometries of Euclidean plane and sphere with radius $R$ that is standardly embedded in 3-dimensional Euclidean space,
can be described by metric tensors of the kind:
\[
g_{E}= dr\otimes dr+r^2d\varphi\otimes d\varphi;\quad
g_{S}=R^2(d\theta\otimes d\theta+\sin^2\theta d\varphi\otimes d\varphi)
\]
for the plane and sphere respectively. Here the pairs $(r,\varphi)$ and $(\theta,\varphi)$ are standard polar coordinates on plane and sphere.
It is well known that these metrics are not isometric, even locally. That answers to the fact that the sphere cannot be mapped onto
plane without deforming the lengths. In the previous section we've established that centrometrics associated with any fixed points on the
plane and on the sphere are equivalent. That can be easily proven. Let us connect selected points with the origins of polar reference frames in
plane and in sphere. Then in these coordinates centrometrics of the plane has the form: $\rho_{E}(r,\varphi)=r,$  and the sphere:
$\rho_{S}(\theta,\varphi)=R\theta.$
If we define the coordinate $r$ (of length) according to formula: $r=R\theta,$  it becomes apparent that
{\it centrometrics of the plane and sphere are indistinguishable from each other.}
Isometries of both centrometrics are infinite-dimensional (isometries of their respective metrics are 3-dimensional, those with a fixed point are
1-dimensional). They can be described by mappings: $\varphi\to
f(\varphi),$ where $f$ is an arbitrary diffeomorphism of a circle $S^1.$

\subsection{ Centrogeometries of Euclidean and pseudo-Euclidean planes}

\begin{figure}[htb]
\centering \unitlength=0.50mm \special{em:linewidth 0.4pt}
\linethickness{0.4pt} \footnotesize \unitlength=0.70mm \special{em:linewidth
0.4pt} \linethickness{0.4pt}
\unitlength=1mm
\special{em:linewidth 0.4pt}
\linethickness{0.4pt}
\begin{picture}(42.17,40.50)
\emline{20.00}{0.33}{1}{20.00}{40.00}{2}
\emline{20.00}{40.00}{3}{19.33}{36.17}{4}
\emline{20.67}{36.17}{5}{20.00}{40.17}{6}
\emline{0.67}{20.00}{7}{40.00}{20.00}{8}
\emline{40.00}{20.00}{9}{35.33}{20.67}{10}
\emline{35.33}{19.33}{11}{40.17}{20.00}{12}
\put(20.00,20.00){\circle{14.00}}
\emline{0.50}{0.50}{13}{39.50}{39.50}{14}
\emline{39.33}{0.33}{15}{0.33}{39.33}{16}
\put(20.00,20.00){\circle{7.36}}
\bezier{200}(7.33,33.17)(20.00,21.17)(32.33,33.17)
\bezier{200}(12.50,27.67)(20.00,20.00)(27.17,27.67)
\bezier{200}(8.33,31.17)(20.00,40.50)(31.17,31.17)
\bezier{200}(8.33,8.50)(20.00,0.00)(31.33,8.33)
\bezier{200}(8.33,31.17)(0.00,20.00)(8.33,8.33)
\bezier{200}(31.17,31.17)(42.17,19.17)(31.17,8.50)
\bezier{200}(6.83,39.83)(20.00,31.83)(32.00,39.83)
\put(21.00,38.17){\makebox(0,0)[lc]{$x_1$}}
\put(38.50,21.00){\makebox(0,0)[lb]{$x_2$}}
\emline{20.00}{20.00}{17}{31.00}{39.33}{18}
\put(27.67,33.50){\circle*{1.33}}
\put(30.83,39.17){\circle*{1.37}}
\bezier{28}(20.00,32.17)(23.83,32.50)(25.83,30.50)
\put(22.50,29.83){\makebox(0,0)[cc]{$\varphi$}}
\end{picture}
\caption{\small Metric circles of Euclidean and pseudo-Euclidean plane}\label{ex2}
\end{figure}
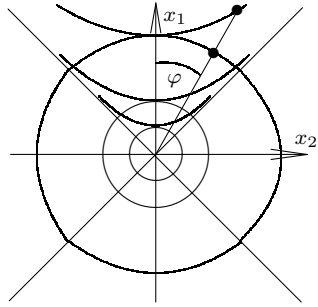

A more interesting example of equivalency between centrometrics can be found in equivalency between Euclidean and pseudo-Euclidean planes.
Fig. \ref{ex2} shows metric sphere families of Euclidean centrometrics: $\rho_E=x_1^2+x_2^2$ and pseudo-Euclidean centrometrics: $\rho_M=x^2_1-x_2^2.$
Let us construct an obvious deformation $\phi:$  ${R^2}\to R^2,$ that implements the equanimity:
\[
\rho_E=\rho_M\circ\phi.
\]
Note that our construction will be of local character, with relation to the upper part of the characteristic cone of centrometric $\rho_M$.
Let us examine metric circles of centrometrics $\rho_E$ and $\rho_M$ with the same radius $r.$
Points of Euclidean circles are described by parametric equations:
\[
x_1=r\cos\varphi;\quad x_2=r\sin\varphi,
\]
and points of pseudo-Euclidean circle by equations:
\[
y_1=r\cosh\psi;\quad y_2=r\sinh\psi,
\]
where $\varphi$ is Euclidean angle, $\psi$  is pseudo-Euclidean angle, $r=\sqrt{x_1^2+x_2^2}$ (in interpretation $\rho_E$)
is  an Euclidean distance from coordinate system  origin, or $r=\sqrt{y_1^2-y_2^2}$ (in interpretation $\rho_M$) is a pseudo-Euclidean
distance to coordinate system origin. In order to construct the map $\phi$ in accordance with the theorem,
we should interpret coordinates $y$ in the terms of Euclidean coordinates $x.$
Euclidean distance to coordinate system origin  for the points of pseudo-Euclidean circle is:
\begin{equation}\label{dist}
r'=r\sqrt{\cosh^2\psi+\sinh^2\psi},
\end{equation}
and coordinates of the circle points with the same radius that lie on the same ray (they're the ones that turn into each other under a deformation sought for)
meet the conditions for affine collinearity:
\[
r\cos\varphi=\lambda r\cosh\psi;\quad r\sin\varphi=\lambda
r\sinh\psi.
\]
By excluding $\lambda$ and $r,$ from latter equanimities we arrive at a condition for angles: $\tan\varphi=\tanh\psi.$
Substituting this condition in (\ref{dist}), after some transformations we get the expression:
\begin{equation}\label{dist1}
r'=r\sqrt{\frac{1+\tan^2\varphi}{1-\tan^2\varphi}}.
\end{equation}
This expression inverts the points of a part of Euclidean circle with radius $r,$ enclosed in the span of angles from $-\pi/4$ to $\pi/4,$
into points of a pseudo-Euclidean circle lying in the upper part of the characteristic cone. Using it, we can construct the map of the
deformation:
\begin{equation}\label{phi}
\phi:\ (r,\varphi)\mapsto(r',\varphi).
\end{equation}
Direct check shows that formula $\rho_E=\rho_M\circ\phi$ takes the form of the equalities:
\[
x_1^2+x_2^2=\stackrel{\phi}{\dots}=y_1^2-y_2^2,
\]
where the Cartesian coordinates are easily mapped with the help of mapping (\ref{phi}):
\[
y_1=x_1\sqrt{\frac{x_1^2+x_2^2}{x_1^2-x_2^2}};\quad
y_2=x_2\sqrt{\frac{x_1^2+x_2^2}{x_1^2-x_2^2}}.
\]
Let us turn to isometry deformations. Using formula (\ref{deff}) in the case $\phi,$  set according to formula (\ref{phi}), and isometry $\chi,$ set
by standard hyperbolic rotation formulae:
\[
y'_1=\cosh\psi\, y_1+\sinh\psi\, y_2;\quad y'_2=\sinh\psi\, y_1+\cosh\psi\,
y_2,
\]
after simple although somewhat lengthy calculations we obtain a deformed isometry:
\[
x'_1=\frac{x_1+\tanh\psi\,x_2}{\sqrt{1+\tanh^2\psi+4\tanh\psi
x_1x_2/r^2}};\quad x'_2=\frac{x_2+\tanh\psi\,x_1}{\sqrt{1+\tanh^2\psi+4\tanh\psi
x_1x_2/r^2}},
\]
where $r^2=x_1^2+x_2^2.$ By direct check we can ascertain that these transformations are in fact exact isometries of Euclidean centrometrics $\rho_E.$
The obtained deformed isometry is as such a Lie group that is isomorphic to a 1-dimensional busts subgroup of  Lorenz group,
which acts nonlinearly in  a coordinate $(x_1,x_2)$ space.
Nonlinear isometry of centrometrics $\rho_M$ can be calculated analogously. It is obtained according to formula (\ref{deff}),
from isometry of Euclidean rotations of metric $\rho_E:$
\begin{equation}\label{eucrot}
x'_1=\cos\varphi\, x_1-\sin\varphi\, x_2;\quad x'_2=\sin\varphi\, x_1+\cos\varphi\,
x_2.
\end{equation}
Here we come to the following transformation formulae:
\[
y'_1=\frac{y_1-\tan\varphi\,y_2}{\sqrt{1-\tan^2\varphi-4\tan\varphi\,
y_1y_2/r^2}};\quad
y'_2=\frac{y_2+\tan\varphi\,y_1}{\sqrt{1-\tan^2\varphi-4\tan\varphi\,
y_1y_2/r^2}},
\]
where $r^2=y_1^2-y_2^2.$
As in previous case it can be ascertained that these formulae describe exact nonlinear isometries of $\rho_M$ centrometric and form a
Lie group that is isomorphic to a 1-dimensional group of Euclidean rotations.

\subsection{Minkowsky space and Berwald-Moore geometry}

There is another interesting possibility of connecting Euclidean and pseudo-Euclidean planes to centrogeometry which generalication
illustrates deformational  equivalency between centrogeometries of quadratic metrics and of Finslerian metrics. Let us take formula for $\rho_M=y_1^2-y_2^2$
and switch to isotropic coordinates: $\xi_1=y_1+y_2,$ $\xi_2=y_1-y_2.$ In these coordinates pseudo-Euclidean centrometric takes the form:
\begin{equation}\label{pseuis}
\rho_M=\xi_1\xi_2.
\end{equation}
Deformed centrometric:
\[
\rho'_M=f\circ\rho_M\circ\phi
\]
is an Euclidean centrometric of the kind $\rho_E=\eta_1^2+\eta_2^2,$ where
\[
\phi:\ (\xi_1,\xi_2)\mapsto (e^{\eta_1^2}, e^{\eta_2^2});\quad f:\
x\to \ln x.
\]
 Group of hyperbolic rotations in isotropic coordinates: $\xi_1\mapsto
t\xi_1,$ $\xi_2\mapsto \xi_2/t$  gets deformed into nonlinear isometry group of Euclidean metric that has the form:
\begin{equation}\label{defm1}
\eta_1\to\sqrt{\eta_1^2+\ln t};\quad \eta_2\to\sqrt{\eta_2^2-\ln t}.
\end{equation}
Note that Euclidean coordinate $(\eta_1,\eta_2)$ map covers only the positive quadrant of the pseudo-Euclidean $(\xi_1,\xi_2)$ map.
Direct generalization of the example examined above for a case of multiple dimensions results in {\it Berwald-Moore centrometric
$\mathcal{H}_n^p$:}
\begin{equation}\label{BM}
\rho_{BM}=\prod\limits_{i=1}^n\xi_i,
\end{equation}
that belongs to the class of Finslerian  metrics. Deformation of the kind:
\[
\phi:\ \xi_i\mapsto e^{\epsilon_i\eta_i^2}\ (i=1\dots n);\quad f:\ x\mapsto
\ln x,
\]
establishes apparent equivalency of Berwald-Moore centrometric to any of the pseudo-Euclidean spaces of the $\mathcal{M}_{p,q}$ $p+q=n$ type.
At the same time, an isometry group of $\mathcal{H}_n^p$ space, that consists of unimodular dilations of the kind:
\[
\xi_i\to t_i\xi_i,\ (i=1,\dots,n),\ \quad  \prod\limits_{i=1}^n t_i=1
\]
gets deformed into nonlinear isometry group of $\mathcal{M}_{p,q}$ spaces of the following kind:
\[
\eta_i\to \sqrt{\epsilon_i\eta_i^2+\ln t_i},\quad i=1,\dots n.
\]
Isometry group $O(p,q)$ of $\mathcal{M}_{p,q}$ space in its turn gets deformed into nonlinear isometries $\mathcal{H}_n^p.$
For example, Euclidean rotation of the kind (\ref{eucrot}) gets strained into space $\mathcal{H}_n^p$ $(n=2)$  transformations of the kind:
\[
\xi'_1=\xi_1^{\cos^2\varphi}\xi_2^{\sin^2\varphi}e^{\sin2\varphi\sqrt{\ln(\xi_1)\ln(\xi_2)}};\quad
\xi'_2=\xi_2^{\cos^2\varphi}\xi_1^{\sin^2\varphi}e^{-\sin2\varphi\sqrt{\ln(\eta_1)\ln(\eta_2)}}.
\]
Equivalency of this type was studied in work \cite{kok2} in view of different conformal gauges that Berwald-Moore metric allows.

\section{Curve straightening}

Let us consider a curve $\gamma:\ R\to \mathcal{M},$ where $\mathcal{M}$ is space with centrometric $\rho_p.$ It is always possible to find a deformation
$\phi_\gamma$ of the space $\mathcal{M},$ such as would invert curve $\gamma$ into straight line $\ell,$
which under appropriate choice of parameters  we could be able to describe by a set of linear functions $l_i(t)$ of some affine parameter $t.$
Indeed, if $x=\varphi(t)$ is a parametric description of the curve $\gamma,$ then a map of the kind:
\[
\phi_{\gamma}=A\circ\varphi^{-1},
\]
applied to parametrical representation of the curve will lead us to the desired result. Here  $A$ stands for general inhomogeneous
affine transformation of the space $\mathcal{M},$ while $\rho_\gamma=\rho\circ\phi_{\gamma}$ is a deformed
centrometric, where curve $\gamma$ is a straight line. It is apparent that any curves family can be similarly transformed into straight lines family.

Of course, same method can be used in standard geometry. For example, parametrical circle equation in polar reference frame: $\rho=\text{const},\ \varphi=t,$
implies that switching to polar reference frame enables strengthening of circles with the center at coordinate system origin.
However, in geometry we hold a passive view as to coordinate switch: points of the space stay where they were and only
their numeric marks change. In our construction called centrogeometry we hold an active view: coordinate stay the same while
transformations describe deformation of the space itself and, therefore, strain of centrogeometry itself.

\section{Conclusion: centrogeometry and physics
}\label{phys}

Despite its "vague"\, character, concept of centrogeometry is in fact quite up to some situations we can meet in geometry and physics.
Let us turn to the concept of "point observer".  In terms of standard geometry that means an observer whose size (or the size of the
laboratory with all the equipment) is greatly less than distances being measured. If a point observer has a device emitting physical signals
(not only light ones) which can get reflected from objects and come back, and a clock, he is able to construct a centrometric related to the
system of bodies around, after he's introduced an arbitrary local system of "angles". All he'll have to measure is time intervals and apply
some preliminary hypotheses regarding signal propagation laws for the signals which his device emits (similar to Synge's chronogeometry \cite{sing}).
Due to the independence of "longitudinal"\, length scale (centrometric distances) and "transversal"\, angle scale which is being chosen arbitrary,
as mentioned before, and independently of the device emitting signals, "angles"\, system allows for arbitrary deformations which do not change
 the centrometric. "Angles"\, here are used as markers for directions. It is this freedom of choice in regard to angles that gets
 reflected in transformation general form (\ref{gizo}) of centrometric isometry. Strained equivalency of different centrometrics
 reflects the option for changing hypotheses of signal propagation laws that the observer has. So, if "physical environment"\, separating the
 observer from objects is unisotropic, the observer would necessarily formulate a unistropic centrogeometry, the metric spheres of which
 is different from Euclidean spheres. Taking into account spatial dispersion of signal velocity, the observer then would come to strained
 centrometric which can appear to resemble Euclidean ones. Arguably, it is impossible neither to separate the rules of signal propagation from geometry
 nor to introduce the notion of standard two-point metric which would make possible the calculation of distance between two objects way
 too far from the observer. That is, impossible without some additional fundamental reasoning.

At first glance the concept of centrogeometry may seem insufficient for description of {\it observer motions,} when the
 observer moves from one point of space to another. In fact these motions can be described within the scope of centrogeometry if we assume that
{\it  centrometric may depend on time.} The {\it non-static centrogeometry concept} enables us to include both body motions and observer's motions.
 Under certain assumptions as regards physical signal propagation, certain choice of background centrogeometry or geometry based on it
 it becomes possible even to distinguish between observer motions or equivalent body motions and those body motions which cannot be compensated
 for by the observer motions. Therefore, following in the steps of Poincar\'{e} in
 \cite{poinc},
 we can go to the group of "solid body"\, motions, or "Galilean group"\,, which depends heavily on our preliminary assumptions regarding signals
 and can differ from usual Galilean or Lorenz group. At the same time, unlike traditional geometry which is related to the points of
 outer space and moments of outer time, centrogeometry is always related to the observer's perception space. It always {\it
 accompanies}
 the observer and describes succession of "shots"\, in his perception. The notion of standard "outer"\, geometry is rendered secondary by
 this approach; it originates owing to the fact that {\it perception shots of different observers turn out to be coordinated} and this
 coordination can be conveniently described by mathematics, using the language of the universal outer geometry.
 To be sure, all that should be viewed as general considerations suggesting possibility of such an approach.
 Its further development could be the subject for future studies.

A possibility to mathematically "straighten"\, trajectory families, those that can be spotted by observer while watching test body motions
is closely related to the geometrization principle as applied to laws of nature and which we've discussed in the Introduction. Starting with a
 certain initial centrogeometry, the observer could then proceed to use a certain deformed centrogeometry where bodies move along straight-line
 trajectories but may be not uniformly. Then by deforming accordingly the time marks the observer could achieve exactly or approximately uniform
 motion of the bodies along straight trajectories. If the observer were able to subject all the bodies of a certain class to such an approach,
 that would mean he put geometrization principle into practice for the bodies of the class: he'd get rid of "spare"\, forces, "shifting"\,
 their influences to the appropriate centrogeometry. Another interesting possibility arising from trajectory deformations is the possibility
 to switch from one kind of dynamics problems to another, similar the method applied for solving plane problems in Newton classical mechanics
 in the scope of TFCV (complex analysis) \cite{bohl}.

As a conclusion we would like to show still one more situation where the use of centrogeometry can be seen as both objective and necessary,
in a sense.  We mean cosmological observations and models based on them. Owing to the peculiarities of cosmological observations
(under necessary assumptions as to light propagation laws and laws of cosmic objects luminescence), only longitudinal distances are observable.
The switch from "cosmologic centrogeometry"\, to "cosmologic geometry"\, (say, Friedman-Robertson-Walker models)
results in many veritable but not verified  assumptions of physical and geometrical character (which probably cannot be proven).
In this situation any conclusions based on assumptions of certain geometry of the Universe appear relative to a great extent
(see for example \cite{kok3}). In such a situation it would probably be wiser to give up searching "the true geometry"\,
of the Universe and use instead those of its qualities that little, if at all, depend on our geometrical conceptions of it.

\bigskip
\bigskip

Finally I would like to thank D. G. Pavlov to whom I am greatly indebted for productive discussions and financial aid.

\end{document}